\documentclass[10pt]{article}
\usepackage{multicol}
\usepackage{graphicx}
\usepackage{amsmath}
\usepackage[a4paper]{geometry}
\usepackage{hyperref}
\usepackage{rotating}

\begin{document}

\begin{center}
{\Large \bf  Investigation of the freeze-out parameters in B-B, O-O, Ca-Ca and Au-Au collisions at 39 GeV}

\vskip1.0cm

Muhammad~Waqas$^{1,}${\footnote{Email (M.Waqas):
waqas\_phy313@yahoo.com; waqas\_phy313@ucas.ac.cn}}, 
Guang Xiong Peng$^{1,2}$ {\footnote{Corresponding author. Email (G. X. Peng): gxpeng@ucas.ac.cn}},
Fu-Hu Liu$^{3}${\footnote{E-mail: fuhuliu@163.com; fuhuliu@sxu.edu.cn}},
Muhammad Ajaz$^{4}${\footnote{Corresponding author.E-mail: ajaz@awkum.edu.pk; muhammad.ajaz@cern.ch}},
Abd Al Karim Haj Ismail$^{5}${\footnote{E-mail: a.hajismail@ajman.ac.ae}}
\\

{\small\it  $^1$ School of Nuclear Science and Technology, University of Chinese Academy of Sciences,
Beijing 100049, China,

$^2$ Theoretical Physics Center for Science Facilities, Institute of High Energy Physics, Beijing 100049, China,

$^3$ Institute of Theoretical Physics, State key
Laboratory of Quantum Optics and Quantum Optics Devices \& Collaborative Innovation Center of Extreme Optics, Shanxi, Taiyuan, Shanxi 030006, China\\

$^4$ Department of Physics, Abdul Wali Khan University Mardan, Mardan 23200, Pakistan\\

$^5$College of Humanities and Sciences, Ajman University, Ajman, 346, UAE\\

$^6$Nonlinear Dynamics Research Center (NDRC), Ajman University, Ajman, 346, UAE}

\end{center}

\vskip1.0cm

{\bf Abstract:} We analyzed the transverse momentum spectra of proton, deuteron and triton in Boron-Boron (B-B), Oxygen-Oxygen (O-O), and Calcium-Calcium (Ca-Ca) central collisions, as well as in several centrality bins in Gold-Gold (Au-Au) collisions at 39 GeV by using the blast wave model with Tsallis statistics. The bulk
properties in terms of kinetic freeze-out temperature, transverse
flow velocity and kinetic freeze-out volume are extracted from the
model by the least square method. We observed that with increasing the rest mass of the particle, the kinetic
freeze-out temperature becomes larger, while
transverse flow velocity and the kinetic freeze-out volume
reduces. These parameters are also found to depend on
the size of the system. Larger the size of the system, the larger they
are. Furthermore, the kinetic freeze-out temperature in peripheral
Au-Au collisions is close to the central O-O collisions. We also
observed that the above parameters depend on the centrality, and they
decrease from central to peripheral collisions. Besides, we also
extracted the entropy-index parameter $q$, and the parameter $N_0$
which shows the multiplicity. Both of them depend on the size of interacting
the system, rest mass of the particle and centrality. Both $q$ and $N_0$ are
larger for lighter particles, and the former is smaller for large
systems while the latter is larger, and the former decrease with
increasing centrality while the latter increase.
\\



\vskip1.0cm

\begin{multicols}{2}

{\section{Introduction}} The investigation of Quantum
Chromodynamics (QCD) phase diagram is one of the indispensable objective of high energy heavy ion collision experiments \cite{1, 2, 3, 4}.
Normally, the QCD phase diagram is charted as temperature
against baryon chemical potential ($\mu_B$). Let us consider the
creation of a thermalized system in heavy-ion collisions, both temperature
and $\mu_B$ may change with changing the collisions energy \cite{5, 6, 7}
and the size of the interacting system \cite{8, 9}. Theoretically, the
phase diagram undergoes a possible change from a high temperature
and high-density phase which is called Quark-Gluon Plasma (QGP)
phase. The QGP phase is influenced by the partonic degrees of
freedom to a phase where the relevant degrees of freedom are
hadronic \cite{10, 11, 12}. There are many observations that have been
connected with the actuality and existence of a phase with
partonic degrees of freedom in the early stages of heavy-ion
collisions \cite{13, 14, 15, 16, 17, 18}. The well-known examples of such observations
are the suppression of high transverse momentum ($p_T$) particles
production in nucleus-nucleus (AA) collisions relative to scaled
proton-proton collisions \cite{1, 2, 3, 4, 13, 14, 15}, large elliptic flow for the
particles with light as well as heavy (strange) valence quarks and
the dissimilarities between baryons and mesons  at intermediate
$p_T$ in AA collisions \cite{19}.

As an important quantity in the physics of high energy collisions,
the temperature is widely used in different theoretical and
experimental studies. There are different kinds of temperatures in
literature, namely initial temperature, chemical freeze-out
temperature, effective temperature and kinetic freeze-out
temperature. These temperatures occur at different stages of the
evolution system and they are discussed in detail in previous
works \cite{20, 21, 22, 23}. Besides temperature, volume also has an
important role in the high-energy collision processes. Like
temperature, different freeze-out volumes occur at
different stages in system evolution but we will study the final
state particles in this work, therefore we will be limited to the
kinetic freeze-out volume. More details of kinetic freeze-out
volume can be found in our previous studies \cite{24, 25, 26, 27}. It
is very important to extract the freeze-out parameters because
they help give some crucial information about the
final state particles.

The $p_T$ spectra of hadrons
are important tools to understand the dynamics of the
particles production in high energy collisions, and we
can extract the freeze-out parameters such as chemical
freeze-out temperature ($T_{ch}$), kinetic freeze-out temperature
($T_0$), effective temperature ($T$), transverse flow
velocity ($\beta_T$) and kinetic freeze-out volume ($V$) from
the $p_T$ spectra of the particles by using different hydrodynamical
models such as Blastwave model with Boltzmann
Gibbs statisitcs \cite{28, 29, 29a}, blast wave model with
Tsallis statistics \cite{30}, Hagedorn thermal model \cite{31}, Standard
distribution \cite{32}, modified hagedron model \cite{33}
and other thermodynamic models \cite{34, 35, 36, 36a, 36b}.

The blast wave model has been extensively used for
the description of the particle spectra in $AA$ and proton-nucleus ($p-A$) collisions \cite{37, 38, 39, 40, 41, 42}. The present study is performed under the assumption that the blast wave model with  Tsallis statistics (TBW) to be the theoretical framework to analyze $p_T$ spectra of protons, deuterons and tritons in Boron-Boron (B-B), Oxygen--Oxygen (O--O) and Calcium--Calcium (Ca--Ca) central collisions at 39 GeV. We have analyzed the transverse momentum spectra of the above particles as well as anti-deuteron at 39 GeV in different centrality intervals in Gold-Gold (Au--Au) collisions. Different collision systems at the same center-of-mass-energy are taken in order to investigate the dependence of the freeze-out parameters on the size of the system. We used TBW model with linear as well as constant flow profile to scrutinize the difference of the two cases.

It is conjectured that the system experiences a hydrodynamic evolution. However, the fluctuation occurs for the hydrodynamic evolution event by event \cite{43} which will leave footmark on the spectra of the particles in low and intermediate $p_T$ region because of 	incomplete set back by the preceding interaction either at hadronic phase or QGP phase \cite{44, 45, 46}. The emission source distribution of the particle has been changed to Tsallis from Boltzmann distribution to take into consideration the effect of the fluctuations responsible for wide applications of the Tsallis-type of non-extensivity \cite{47}.

The remainder of the paper consists of the method and formalism in section 2, followed by the results
and discussion in section 3. In section 4, we summarized our main observations and conclusions.
\\

{\section{The method and formalism}} With the TBW model \cite{30}, the invariant momentum
distribution is expressed as

\begin{align}
f_1(p_T)=&\frac{1}{N}\frac{\mathrm{d}N}{\mathrm{d}p_\mathrm{T}}=C \frac{gV}{(2\pi)^2} p_T m_T \int_{-\pi}^\pi d\phi\int_0^R rdr \nonumber\\
& \times\bigg\{{1+\frac{q-1}{T_0}} \bigg[m_T  \cosh(\rho)-p_T \sinh(\rho) \nonumber\\
& \times\cos(\phi)\bigg]\bigg\}^\frac{-q}{(q-1)}
\end{align}
where $C$, $g$ and $V$ are the normalized constant, degeneracy factor and kinetic freeze-out volume respectively.
$m_T$ is the transverse mass and is given as $m_T=\sqrt{p_T^2+m_0^2}$ and $m_0$ is the rest mass. $\phi$ is the azimuthal angle, $R$ denotes the maximum $r$ while $r$ is the radial coordinate. $q$ is entropy based parameter and it shows the deviation of the system from equilibrium. $\rho$ is the boost angle and is given as $\rho=\tanh^{-1} [\beta(r)]$, whereas $\beta(r)$ is a self-similar flow profile and is given as $\beta(r)=\beta_S(r/R)^{n_0}$.
$\beta_S$ express the flow velocity on the surface and as mean of $\beta(r)$, one has $\beta_T=(2/R^2)\int_0^R r\beta(r)dr=2\beta_S/(n_0+2)$.

we used eq. (1) for the analysis in the present work, because the $p_T$ range is not so wide and we have used a single component of TBW model. However, in a wider $p_T$ range where hard scattering is involved, one can use the superposition of soft and hard component, or the usual step function. In order to understand the whole methodology, one can read our previous works \cite{48, 48a}.
\\

\section{Results and discussion}

\subsection{Comparison with data}

We analyzed the double differential $p_T$ spectra invariant yield of $p$, $d$ and $t$ in B-B, O-O and Ca-Ca central collisions at the center of mass energy per nucleon $\sqrt{s_{NN}}=$39 GeV. The experimental data are indicated by the symbols and different symbols indicate different particles. The curves over the data are the numerical result of the fit by the blast wave model with Tsallis statistics i.e Eq. (1). The solid curves are the fit results for which the flow is constant ($n_0$=0), and the dashed curves are the results with linear flow ($n_0$=1). It is found that the model results can well describe the experimental data of STAR Collaboration \cite{49}. The corresponding data by fit ratios are given in the lower panel of the figure. The filled and open symbols in the data/fit ratio indicate the deviation of the curves from the data with $n_0$=0 and $n_0$=1 respectively. The extracted values of the relative parameters are listed in table 1 and 2 along with the values of $\chi^2$/$dof$, where $dof$ denoted the number of degrees of freedom. {\bf Large deviation of the fit line from the data is seen, especially at the last points in some cases. This is caused by two reasons. On the one hand, the statistics is low at the last points. If the statistics is high, the situation is expected to be changeable. On the other hand, a two-component function is needed. However, the second component from the high $p_T$ region contributes slightly to the parameters. In fact, we do not needed to consider the second component in the present work.}
\begin{figure*}[htb!]
\begin{center}
\hskip-0.153cm
\includegraphics[width=15cm]{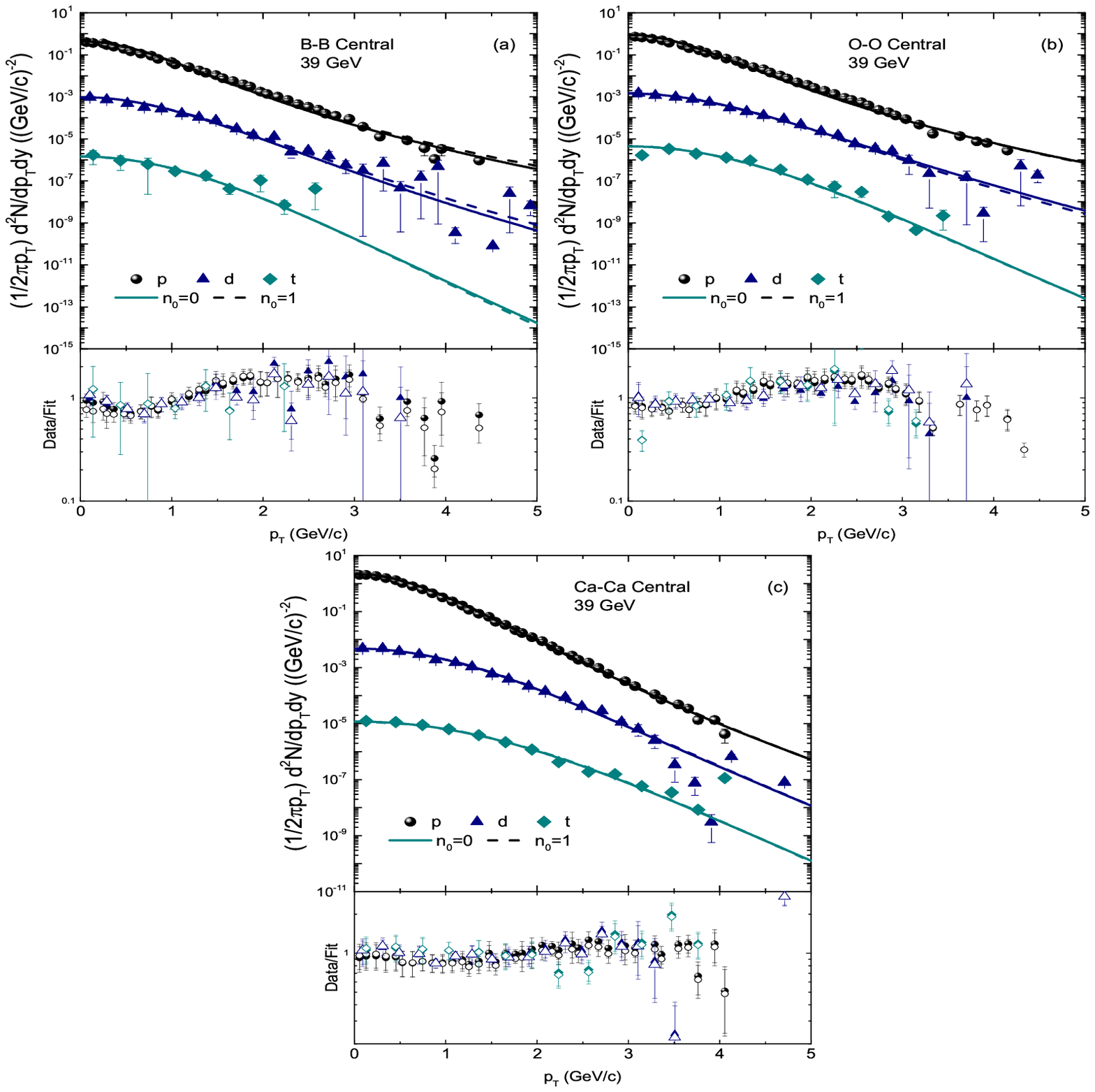}
\end{center}
Fig. 1. Transverse momentum spectra of proton, deuteron and triton produced
in the most central ($0--10\%$) B-B, O-O and Ca-Ca collision at 39 GeV.
The symbols indicate the experimental data of STAR Collaboration \cite{49} and
the curves are our fit results by the blast wave model with Tsallis statistics
with $n_0$=0 and $n_0$=1. The solid and dashed curves are the fit results
with $n_0$=0 and $n_0$=1 respectively. The lower panel of the figure represents
the data by fit ratios.
\end{figure*}

\begin{figure*}[htb!]
\begin{center}
\hskip-0.153cm
\includegraphics[width=15cm]{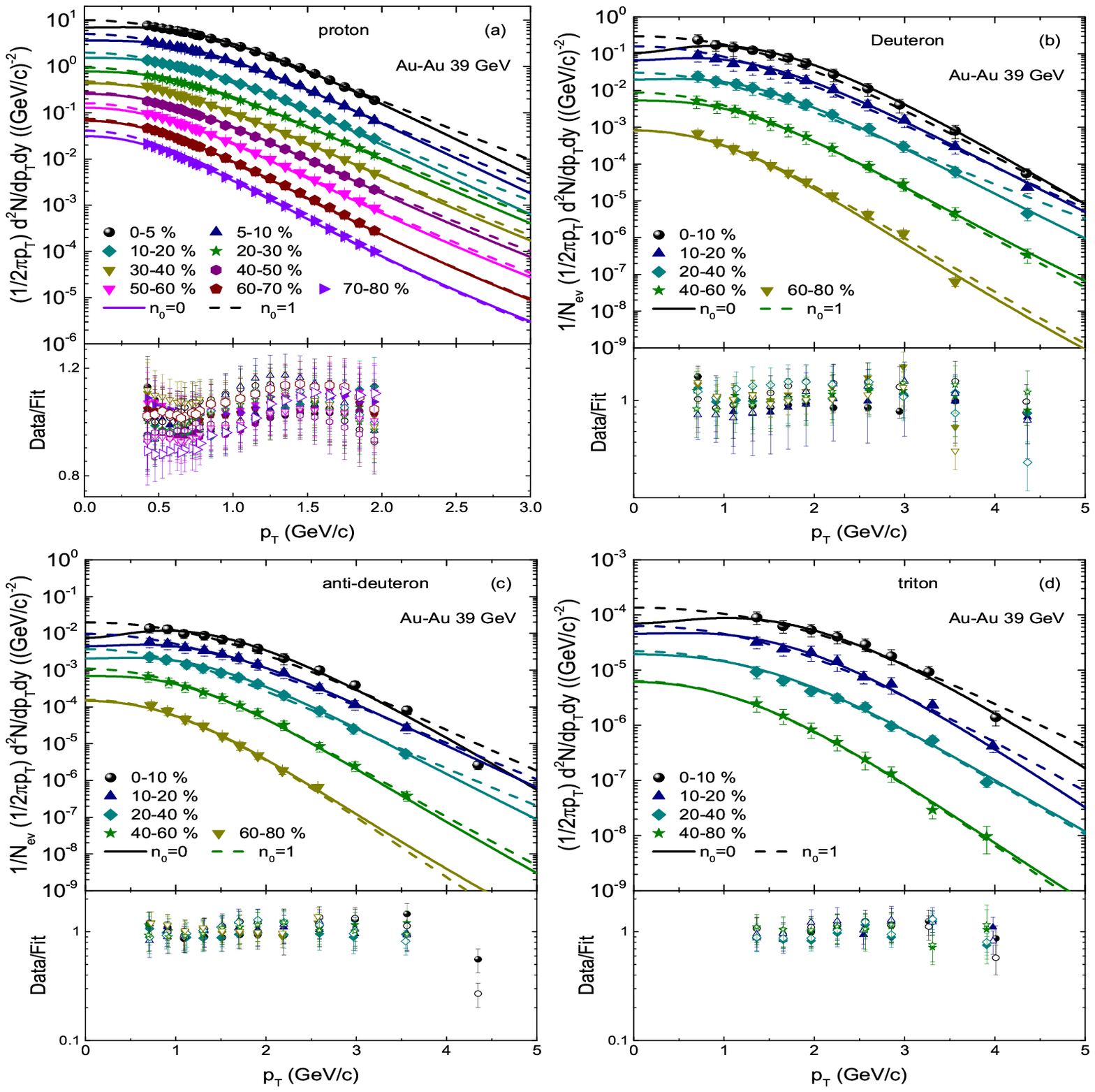}
\end{center}
Fig. 2. Transverse momentum spectra of proton, deuteron,
anti-deuteron and triton are produced in the different centrality
intervals in Au-Au collision at 39 GeV. The symbols indicate the
experimental data \cite{50, 51, 52} of STAR Collaboration and the
curves are our fit results by Eq. (1) with $n_0$=0 and $n_0$=1.
The solid and dashed curves are the fit results with $n_0$=0 and
$n_0$=1 respectively. The lower panel of the figure represents the
data by fit ratios.
\end{figure*}
Figure 2 is similar to fig. 1, but it shows the event centrality
dependent double differential $p_T$ spectra of $p$, $d$, $\bar d$
and $t$ produced in Au-Au collisions at 39 GeV. The experimental
data \cite{50, 51, 52} of the STAR Collaboration is indicated by the
symbols and the curves over the data are our fit results by Eq.
(1). Different symbols in each panel indicate different
centrality intervals. The solid and dashed lines over the data are
the fit results with $n_0$=0 and $n_0$=1 respectively. One can
find that the model can fit the data well. In the lower part of
each panel, the corresponding data by fit ratios are given in
order to show the deviation of the curve from the experimental
data. The filled and open symbols in the lower part of each panel
show the results of data/fit ratios by $n_0$=0 and $n_0$=1
respectively.

\textbf{It is noteworthy that the values in Tables 1 and 2 are
extracted from Eq. (1). The values of $\chi^2$ show the deviation of
the fit line from the data points and dof is the number of degrees
of freedom which is the dof-number of free parameters. $N_0$ is the
normalization constant. The normalization constant $C$
used in Eq. (1) and $N_0$ are not the same. The normalization
constant $C$ is used to let the integral of Eq. (1) be unity,
while $N_0$ is the normalization constant which is used to compare
the fit function $f_S(p_T)$ and the experimental spectra.}

\begin{table*}[tb!]
{\scriptsize Table 1. Values of free parameters $T_0$,
$\beta_T$, q and $V$, normalization constant ($N_0$),
$\chi^2$ and degree of freedom (dof) corresponding to the curves
in Fig. 1--2 with $n_0$=0 \vspace{-.50cm}
\begin{center}
\begin{tabular}{ccccccccccc}\\ \hline\hline
  & Collision &  Centrality   & Particle & $T_0$ (GeV)     & $\beta_T$ (c)    & $V (fm^3)$    & $q$             & $N_0$          & $\chi^2$/ dof\\\hline
  & B-B       & $0-10\%$   & $p$  &$0.035\pm0.005$  & $0.196\pm0.011$  & $2100\pm100$  & $1.085\pm0.005$  &$0.095\pm0.005$             & 138/33\\
   & --       & $0-10\%$    & $d$  &$0.045\pm0.005$  & $0.140\pm0.008$  & $1720\pm104$ & $1.050\pm0.004$    &$2.7\times10^{-4}\pm5\times10^{-5}$&333.2/20\\
   & --       & $0-10\%$    & $t$  &$0.053\pm0.005$  & $0.080\pm0.009$  & $1413\pm107$  & $1.023\pm0.007$  &$8\times10^{-7}\pm4\times10^{-8}$  &5.1/5\\
 \hline
 & O-O       & $0-10\%$  & $p$    &$0.044\pm0.004$  & $0.232\pm0.011$  & $2645\pm108$  & $1.084\pm0.006$   &$0.14\pm0.05$                    & 228.4/34\\
 &           & $0-10\%$  & $d$    &$0.056\pm0.006$  & $0.173\pm0.010$  & $2376\pm120$  & $1.050\pm0.007$  &$3.6\times10^{-4}\pm5\times10^{-5}$ &236/17\\
 &           & $0-10\%$  & $t$    &$0.062\pm0.004$  & $0.115\pm0.004$  & $2000\pm100$  & $1.022\pm0.004$  &$2\times10^{-6}\pm4\times10^{-7}$  & 66.5/8\\
 \hline
 & Ca-Ca     & $0-10\%$  & $p$    &$0.083\pm0.005$  & $0.313\pm0.010$  & $3287\pm115$  & $1.042\pm0.006$  &$0.36\pm0.05$                       & 34.4/38\\
 & --        & $0-10\%$  & $d$    &$0.092\pm0.005 $ & $0.261\pm0.012$  & $3000\pm120$  & $1.028\pm0.003$  &$0.0012\pm0.0005$                   & 63.1/18\\
 & --        & $0-10\%$  & $t$    &$0.104\pm0.005$  & $0.216\pm0.010$  & $2756\pm120$  & $1.018\pm0.004$  &$7\times10^{-6}\pm5\times10^{-7}$   & 34.3/10\\
 \hline
   & Au-Au   & $0-5\%$   & $p$    &$0.108\pm0.005$  & $0.320\pm0.010$  & $5600\pm200$  & $1.020\pm0.003$   &$1.45\pm0.08$                       & 3.4/18\\
 & --        & $5-10\%$  & $p$    &$0.100\pm0.005$  & $0.300\pm0.010$  & $5100\pm160$  & $1.030\pm0.004$  &$0.7\pm0.05$                         & 3/18\\
 & --        & $10-20\%$ & $p$    &$0.093\pm0.005$  & $0.290\pm0.007$  & $4618\pm148$  & $1.036\pm0.006$  &$0.3\pm0.04$                        & 11.2/18\\
 & --        & $20-30\%$ & $p$    &$0.086\pm0.005$  & $0.274\pm0.009$  & $4000\pm123$  & $1.050\pm0.005$  &$0.16\pm0.03$                        & 6.2/18\\
 & --        & $30-40\%$ & $p$    &$0.079\pm0.005$  & $0.250\pm0.011$  & $3400\pm200$  & $1.057\pm0.005$  &$0.092\pm0.006$                     & 2.3/18\\
 & --        & $40-50\%$ & $p$    &$0.072\pm0.005$  & $0.233\pm0.010$  & $3219\pm130$  & $1.064\pm0.005$  &$0.048\pm0.004$                     & 4/18\\
 & --        & $50-60\%$ & $p$    &$0.061\pm0.005$  & $0.219\pm0.010$  & $2900\pm140$  & $1.071\pm0.005$  &$0.025\pm0.005$                    & 7.4/18\\
 & --        & $60-70\%$ & $p$    &$0.051\pm0.005$  & $0.200\pm0.010$  & $2749\pm126$  & $1.076\pm0.004$  &$0.012\pm0.005$                   & 4.1/18\\
 & --        & $70-80\%$ & $p$    &$0.042\pm0.005$  & $0.183\pm0.009$  & $2300\pm160$  & $1.082\pm0.004$  &$0.006\pm0.0003$                   & 5.5/18\\
  \hline
   & Au-Au   & $0-10\%$ & $d$   &$0.122\pm0.005$  & $0.321\pm0.011$  & $4600\pm210$  & $1.010\pm0.003$   &$0.08\pm0.004$                      & 2.5/8\\
 &  --       & $10-20\%$  & $d$ &$0.106\pm0.006$  & $0.290\pm0.011$  & $4158\pm183$  & $1.025\pm0.004$  &$0.035\pm0.005$                      & 1.9/8\\
 &  --       & $20-40\%$ & $d$  &$0.093\pm0.005$  & $0.270\pm0.010$  & $3500\pm155$  & $1.032\pm0.005$   &$0.0098\pm0.0006$                    & 0.8/8\\
 &  --       & $40-60\%$ & $d$  &$0.078\pm0.006$  & $0.227\pm0.013$  & $3320\pm131$  & $1.038\pm0.005$  &$0.0018\pm0.0004$                    & 1.1/8\\
 &  --       & $60-80\%$ & $d$  &$0.060\pm0.005$  & $0.176\pm0.010$  & $2900\pm180$  & $1.040\pm0.004$  &$2\times10^{-4}\pm4\times10^{-5}$    & 9.9/7\\
\hline
  & Au-Au    & $0-10\%$ & $\bar d$   &$0.122\pm0.005$  & $0.320\pm0.010$  & $4610\pm182$  & $1.01\pm0.004$   &$0.0058\pm0.0005$              & 13.8/8\\
  &  --      & $10-20\%$  & $\bar d$ &$0.107\pm0.004$  & $0.290\pm0.010$  & $4158\pm160$  & $1.044\pm0.005$   &$0.0023\pm0.0004$              & 0.5/8\\
  &  --      & $20-40\%$ & $\bar d$  &$0.093\pm0.005$  & $0.270\pm0.008$  & $3500\pm162$  & $1.031\pm0.004$   &$0.0011\pm0.0003$              & 1/8\\
  &  --      & $40-60\%$ & $\bar d$  &$0.078\pm0.005$  & $0.227\pm0.009$  & $3335\pm140$  & $1.032\pm0.004$  &$2.5\times10^{-4}\pm5\times10^{-5}$ &1.5/7\\
  &  --      & $60-80\%$ & $\bar d$  &$0.060\pm0.005$  & $0.176\pm0.010$  & $2889\pm100$  & $1.051\pm0.005$  &$3.8\times10^{-5}\pm5\times10^{-6}$ &1.88/5\\
\hline
   & Au-Au   & $0-10\%$ & $t$   &$0.129\pm0.004$  & $0.280\pm0.011$  & $3400\pm160$  & $1.020\pm0.003$   &$1.4\times10^{-4}\pm4\times10^{-5}$ & 0.95/4\\
   & --      & $10-20\%$ & $t$  &$0.115\pm0.005$  & $0.242\pm0.010$  & $3130\pm133$  & $1.025\pm0.005$  &$6.1\times10^{-5}\pm4\times10^{-6}$  & 1.98/4\\
   & --      & $20-40\%$ & $t$  &$0.88\pm0.005$  & $0.200\pm0.011$  & $2800\pm150$  & $1.044\pm0.005$  &$2\times10^{-5}\pm5\times10^{-6}$   & 7.2/4\\
   & --      & $40-80\%$ & $t$  &$0.068\pm0.004$  & $0.150\pm0.008$  & $2600\pm103$  & $1.045\pm0.004$  &$4.5\times10^{-6}\pm7\times10^{-7}$  & 2.2/4\\
  \hline
\end{tabular}%
\end{center}}
\end{table*}

\begin{table*}[tb!]
{\scriptsize Table 2. Values of free parameters $T_0$,
$\beta_T$, q and $V$, normalization constant ($N_0$),
$\chi^2$ and degree of freedom (dof) corresponding to the curves
in Fig. 1--2 with $n_0$=1. \vspace{-.50cm}
\begin{center}
\begin{tabular}{ccccccccccc}\\ \hline\hline
   & Collision &  Centrality   & Particle & $T_0$ (GeV)     & $\beta_T$ (c)    & $V (fm^3)$    & $q$             & $N_0$          & $\chi^2$/ dof\\\hline
   & B-B       & $0-10\%$     & $p$      &$0.040\pm0.004$  & $0.200\pm0.009$  & $2100\pm92$   & $1.090\pm0.004$  &$0.1\pm0.03$              & 239.7/33\\
   & --       & $0-10\%$      & $d$     &$0.052\pm0.005$  & $0.151\pm0.010$  & $1720\pm104$  & $1.050\pm0.004$ &$2.7\times10^{-4}\pm5\times10^{-5}$&311.3/20\\
   & --       & $0-10\%$      & $t$     &$0.060\pm0.006$  & $0.090\pm0.008$  & $1413\pm107$  & $1.020\pm0.007$ &$8\times10^{-7}\pm4\times10^{-8}$  &5/5\\
     \hline
   & O-O       & $0-10\%$     & $p$  &$0.048\pm0.005$  & $0.236\pm0.010$  & $2645\pm108$  & $1.080\pm0.006$   &$0.14\pm0.05$                    & 212.2/34\\
   & --        & $0-10\%$  & $d$    &$0.061\pm0.005$  & $0.180\pm0.011$  & $2376\pm120$  & $1.050\pm0.007$  &$3.6\times10^{-4}\pm5\times10^{-5}$ &209.5/17\\
   & --        & $0-10\%$  & $t$    &$0.069\pm0.006$  & $0.117\pm0.009$  & $2000\pm100$  & $1.020\pm0.004$  &$2\times10^{-6}\pm4\times10^{-7}$   & 65.5/8\\
   \hline
    & Ca-Ca     & $0-10\%$  & $p$    &$0.088\pm0.006$  & $0.322\pm0.010$  & $3287\pm115$  & $1.038\pm0.006$  &$0.36\pm0.05$                       & 36.4/38\\
    & --        & $0-10\%$  & $d$    &$0.097\pm0.005 $ & $0.270\pm0.011$  & $3000\pm120$  & $1.025\pm0.003$  &$0.0012\pm0.0005$                   & 58.7/18\\
    & --        & $0-10\%$  & $t$    &$0.110\pm0.004$  & $0.225\pm0.012$  & $2756\pm120$  & $1.015\pm0.004$  &$7\times10^{-6}\pm5\times10^{-7}$   & 35.4/10\\
  \hline
    & Au-Au     & $0-5\%$   & $p$    &$0.118\pm0.006$  & $0.380\pm0.013$  & $5600\pm200$  & $1.030\pm0.005$   &$1.45\pm0.08$                        & 13/18\\
    & --        & $5-10\%$  & $p$    &$0.110\pm0.005$  & $0.350\pm0.010$  & $5100\pm160$  & $1.037\pm0.003$  &$0.7\pm0.05$                        & 20.2/18\\
    & --        & $10-20\%$ & $p$    &$0.102\pm0.005$  & $0.320\pm0.007$  & $4618\pm148$  & $1.047\pm0.006$  &$0.3\pm0.04$                        & 9.7/18\\
    & --        & $20-30\%$ & $p$    &$0.096\pm0.005$  & $0.330\pm0.008$  & $4000\pm123$  & $1.055\pm0.005$  &$0.16\pm0.03$                       & 4.8/18\\
    & --        & $30-40\%$ & $p$    &$0.089\pm0.005$  & $0.320\pm0.009$  & $3400\pm200$  & $1.057\pm0.004$  &$0.088\pm0.006$                     & 11.9/18\\
    & --        & $40-50\%$ & $p$    &$0.080\pm0.006$  & $0.300\pm0.012$  & $3219\pm130$  & $1.064\pm0.005$  &$0.05\pm0.004$                       & 4.9/18\\
    & --        & $50-60\%$ & $p$    &$0.068\pm0.004$  & $0.267\pm0.014$  & $2900\pm140$  & $1.071\pm0.006$  &$0.028\pm0.004$                    & 11.88/18\\
    &           & $60-70\%$ & $p$    &$0.056\pm0.007$  & $0.246\pm0.013$  & $2749\pm126$  & $1.076\pm0.004$  &$0.012\pm0.005$                   & 11.3/18\\
    &           & $70-80\%$ & $p$    &$0.045\pm0.005$  & $0.217\pm0.010$  & $2300\pm160$  & $1.080\pm0.005$  &$0.007\pm0.0005$                   & 32.4/18\\
  \hline
   & Au-Au     & $0-10\%$ & $d$   &$0.130\pm0.005$  & $0.358\pm0.009$  & $4600\pm210$  & $1.010\pm0.003$   &$0.08\pm0.004$                      & 3/8\\
   & --        & $10-20\%$  & $d$ &$0.114\pm0.006$  & $0.308\pm0.011$  & $4158\pm183$  & $1.030\pm0.004$   &$0.04\pm0.005$                      & 5.2/8\\
   & --        & $20-40\%$ & $d$  &$0.102\pm0.005$  & $0.279\pm0.009$  & $3500\pm155$  & $1.050\pm0.004$   &$0.009\pm0.0004$                    & 26.6/8\\
   & --        & $40-60\%$ & $d$  &$0.090\pm0.006$  & $0.269\pm0.011$  & $3320\pm131$  & $1.030\pm0.005$  &$0.0022\pm0.0003$                    & 2/8\\
   & --        & $60-80\%$ & $d$  &$0.083\pm0.005$  & $0.255\pm0.012$  & $2900\pm180$  & $1.030\pm0.005$  &$2\times10^{-4}\pm4\times10^{-5}$    & 25/7\\
\hline
   & Au-Au     & $0-10\%$ & $\bar d$   &$0.130\pm0.004$  & $0.358\pm0.007$  & $4592\pm184$  & $1.020\pm0.003$   &$0.0065\pm0.0005$              &51.8/8\\
   & --        & $10-20\%$  & $\bar d$ &$0.114\pm0.005$  & $0.308\pm0.010$  & $4150\pm160$  & $1.040\pm0.005$   &$0.0029\pm0.0005$              & 2.1/8\\
   & --        & $20-40\%$ & $\bar d$  &$0.100\pm0.004$  & $0.274\pm0.009$  & $3558\pm162$  & $1.045\pm0.005$   &$0.0011\pm0.0003$              & 3.1/8\\
   & --        & $40-60\%$ & $\bar d$  &$0.092\pm0.005$  & $0.270\pm0.012$  & $3317\pm140$  & $1.030\pm0.004$   &$2.5\times10^{-4}\pm5\times10^{-5}$ &0.95/7\\
   & --        & $60-80\%$ & $\bar d$  &$0.081\pm0.005$  & $0.250\pm0.008$  & $2904\pm100$  & $1.032\pm0.005$   &$3.8\times10^{-5}\pm5\times10^{-6}$& 2.9/5\\
\hline
   & Au-Au     & $0-10\%$ & $t$   &$0.139\pm0.006$  & $0.340\pm0.010$  & $3400\pm160$  & $1.030\pm0.003$   &$1.5\times10^{-4}\pm5\times10^{-5}$ & 6.8/4\\
   & --        & $10-20\%$ & $t$  &$0.123\pm0.004$  & $0.300\pm0.012$  & $3130\pm133$  & $1.031\pm0.004$  &$6\times10^{-5}\pm4\times10^{-6}$   & 1.8/4\\
   & --        & $20-40\%$ & $t$  &$0.114\pm0.005$  & $0.271\pm0.010$  & $2800\pm150$  & $1.032\pm0.005$  &$2\times10^{-5}\pm5\times10^{-6}$   & 5/4\\
   & --        & $40-80\%$ & $d$  &$0.090\pm0.005$  & $0.216\pm0.010$  & $2600\pm103$  & $1.033\pm0.004$  &$4.5\times10^{-6}\pm7\times10^{-7}$  & 1/4\\
  \hline
\end{tabular}%
\end{center}}
\end{table*}

\subsection{Parameters trend}
To investigate the trend of parameters with the rest mass $m_0$ of
the particle and with centrality, we present Figure 3. Panel (a)
presents the dependence of $T_0$ on $m_0$, while panel (b) shows
the rest mass as well as centrality dependence of $T_0$. Panel (a)
exhibits the result of $T_0$ in B-B, O-O and Ca-Ca collisions.
{\bf The legends of each panel are shown in their corresponding
right panel}. {\bf Each collision system is represented by
different symbols and the solid and open symbols represent the
results from the TBW model with $n_0$=0 and $n_0$=1 respectively in
panel (a), however in panel (b) each symbol represents different
particles and the filled (upper half-filled symbol) and the empty
(lower half-filled symbol) show the results for $n_0$=0 and
$n_0$=1 respectively}. We noticed that $T_0$ increased with $m_0$
which shows the mass differential freeze-out scenario where
the heavier particles freeze-out early, and it is inconsistent
with our previous results \cite{20, 22, 24, 25, 26}. Furthermore, $T_0$
in Ca-Ca collisions is larger than in O-O collisions and in the
latter, it is larger than in B-B collisions. In panel (b) the dependence of $T_0$ on the centrality of collision is shown. Larger $T_0$ in a central collision is observed due to the involvement of large number of participants in the central collisions where the collision is more
harsh and more energy is deposited per nucleon, however as we go
towards the periphery, the participants become less which leads to
less violent collision and that results in a transfer of less amount
of energy per nucleon and consequently $T_0$ decreases. {\bf It is
noteworthy that the temperature refers to the final state
observable, and it might also be lower if the particles decouple
later in central collisions. That is to say, whether the
temperature is higher or lower in central collisions, we have a
suitable explanation. The present work shows that the temperature
is higher in central collisions. We explain the higher temperature
as a larger energy release.}

Like in panel (a), $T_0$ increases with $m_0$ in Au-Au collisions
too. $T_0$ in Au-Au collision is larger than in B-B, O-O and Ca-Ca
collisions which demonstrate its dependence on the size of the
system. The larger the system size, the larger is the kinetic freeze-out
temperature because a large system contains a large number of nucleons
and the collision among large system is very harsh which highly
increase the degree of excitation of the system compared to the
small systems. We further noticed that the values of $T_0$ in the
most peripheral Au-Au collisions are close to the central O-O
collisions. For instance in table 1, the values $T_0$ for $p$, $d$, and
$t$ in Au-Au collisions are $0.042\pm0.005$,
$0.060\pm0.005$ and $0.068\pm0.004$ respectively, while in case of O-O most central collisions, their values are
$0.044\pm0.004$, $0.056\pm0.006$ and $0.062\pm0.004$ respectively.
This indicates that the central O-O collisions and peripheral
Au-Au collisions at the same center-of-mass energy have similar
thermodynamic nature. Similarly in Table 2, $T_0$ for $p$, $d$ and
$t$ emissions in Au-Au peripheral collision is $0.045\pm0.005$,
$0.083\pm0.005$ and $0.090\pm0.005$ respectively and for O-O
central collisions they are $0.048\pm0.005$, $0.061\pm0.005$ and
$0.069\pm0.005$ respectively. In this case, p in central O-O and
peripheral Au-Au shows similar thermodynamic nature, and deuteron
and triton show close value. In addition, $T_0$ is observed to be
larger at $n_0$=1 than $n_0$=0.

\begin{figure*}[htb!]
\begin{center}
\hskip-0.153cm
\includegraphics[width=15cm]{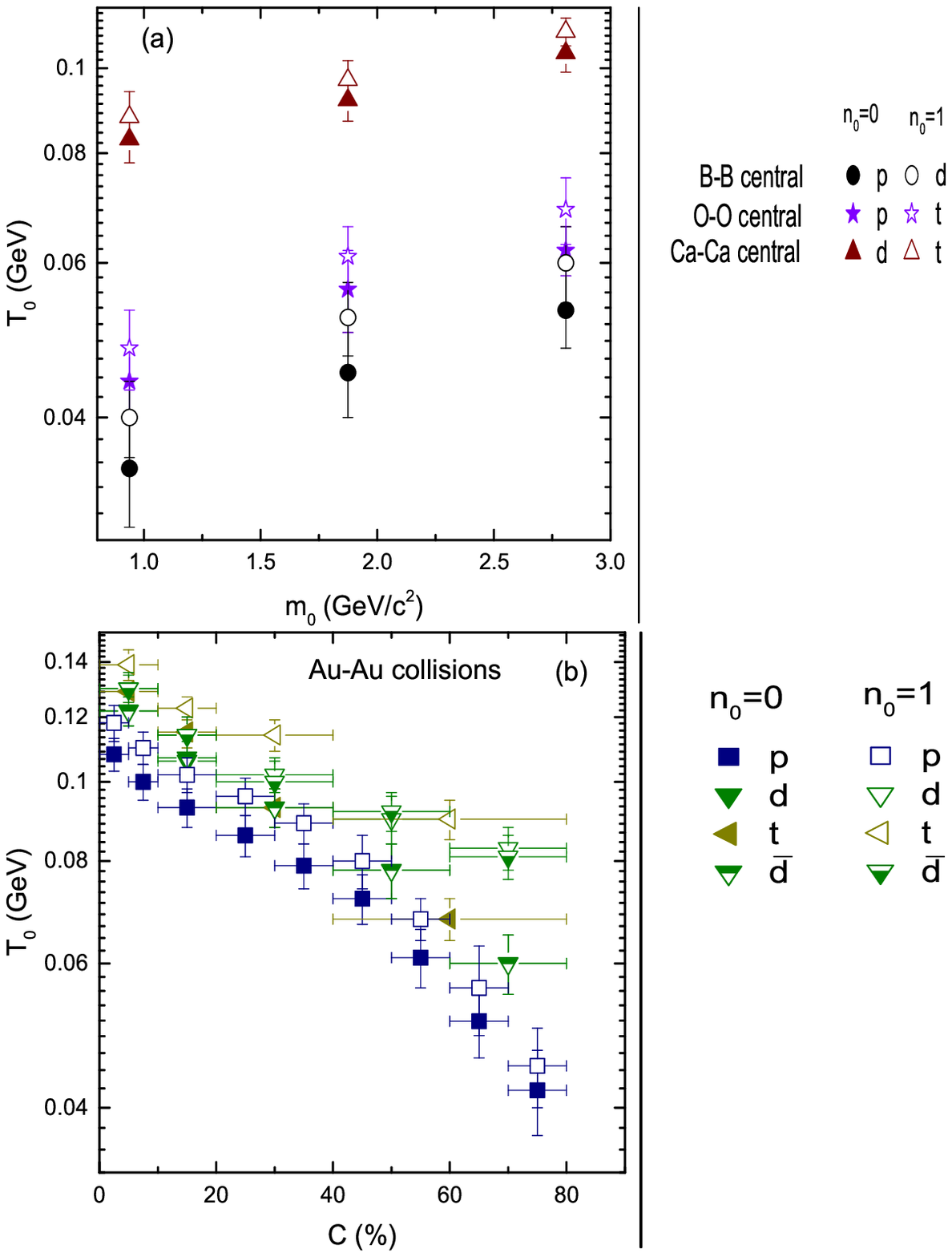}
\end{center}
Fig. 3. Dependence of (a) $T_0$ on $m_0$ and (b) $T_0$ on $m_0$ as well as on centrality.
\end{figure*}

Fig. 4 is similar to Fig. 3, but it demonstrates the dependence of
$\beta_T$ on $m_0$ in panel (a) and on centrality and $m_0$ in
panel (b). It is investigated that $\beta_T$ depends on $m_0$.
The larger the mass of the particle, the smaller is $\beta_T$ values. In panel
(b), $\beta_T$ is dependent on the size of the interacting system.
The larger the size of the system is, the larger the $\beta_T$. The
collisions of large systems are pretty violent which deposits more
energy in the system and it expands quickly. In panel (b)
$\beta_T$ gives an increasing trend with increasing centrality due
to the fact that the collisions become more violent as the system
goes toward centrality and the deposition of energy is large which
results in the quick expansion of the fireball. Likewise $T_0$,
$\beta_T$ is also larger with $n_0$=1 in TBW model.
\begin{figure*}[htb!]
\begin{center}
\hskip-0.153cm
\includegraphics[width=15cm]{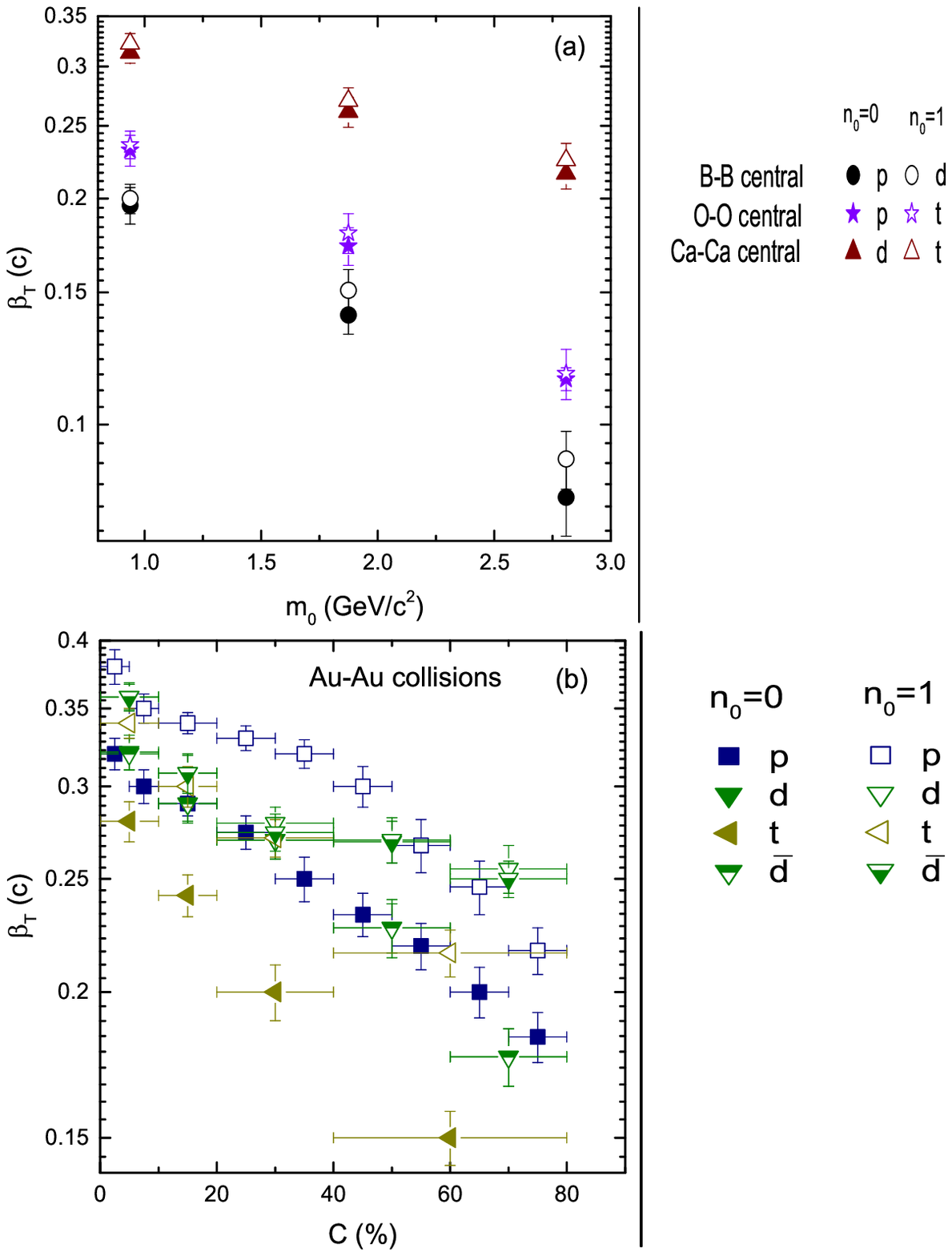}
\end{center}
Fig. 4. Dependence of (a) $\beta_T$ on $m_0$ (b) $T_0$ on $m_0$ as well as on centrality.
\end{figure*}
\begin{figure*}[htb!]
\begin{center}
\hskip-0.153cm
\includegraphics[width=15cm]{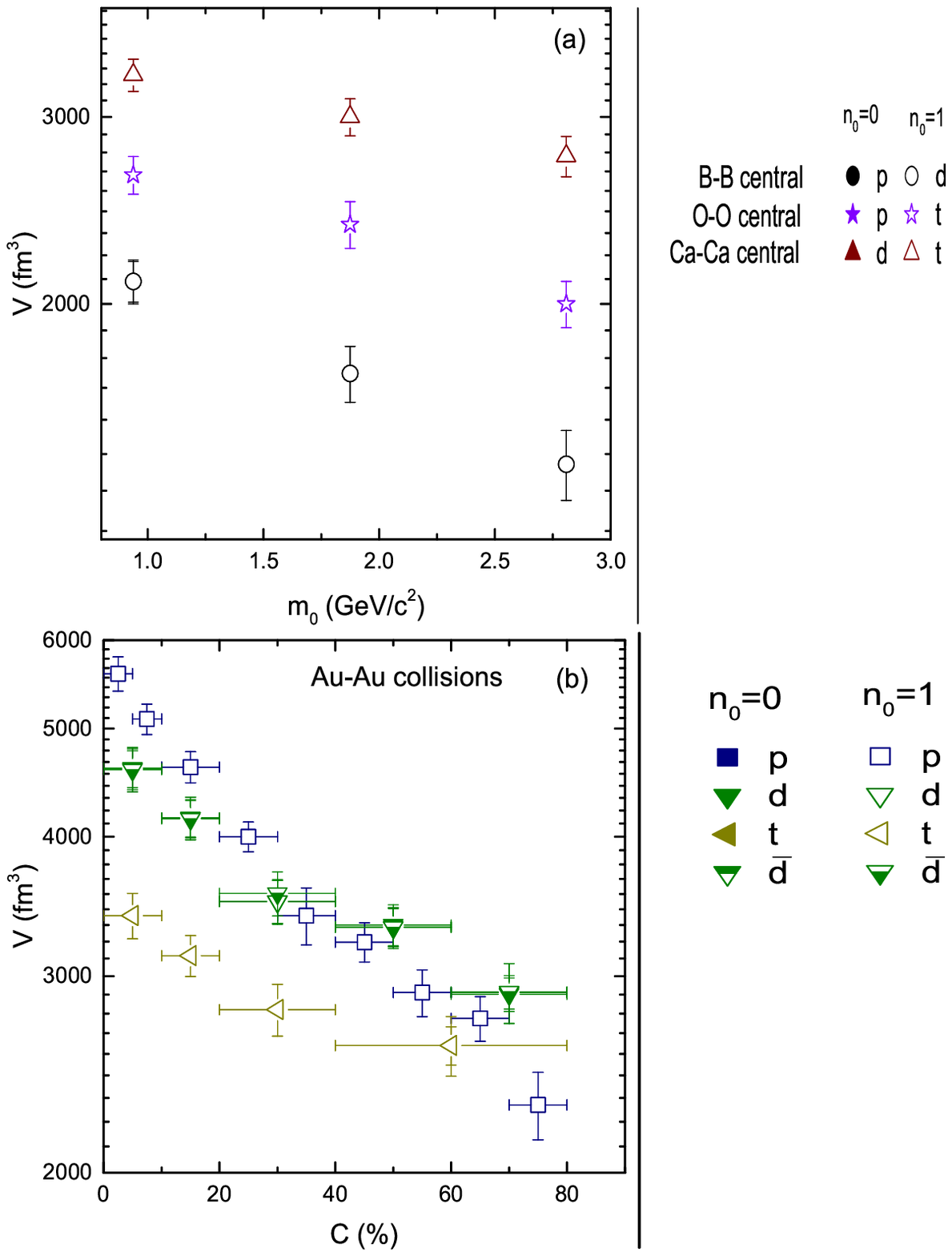}
\end{center}
Fig. 5. Dependence of (a) $V$ on $m_0$ (b) $T_0$ on $m_0$ as well as on centrality.
\end{figure*}

\begin{figure*}[htb!]
\begin{center}
\hskip-0.153cm
\includegraphics[width=15cm]{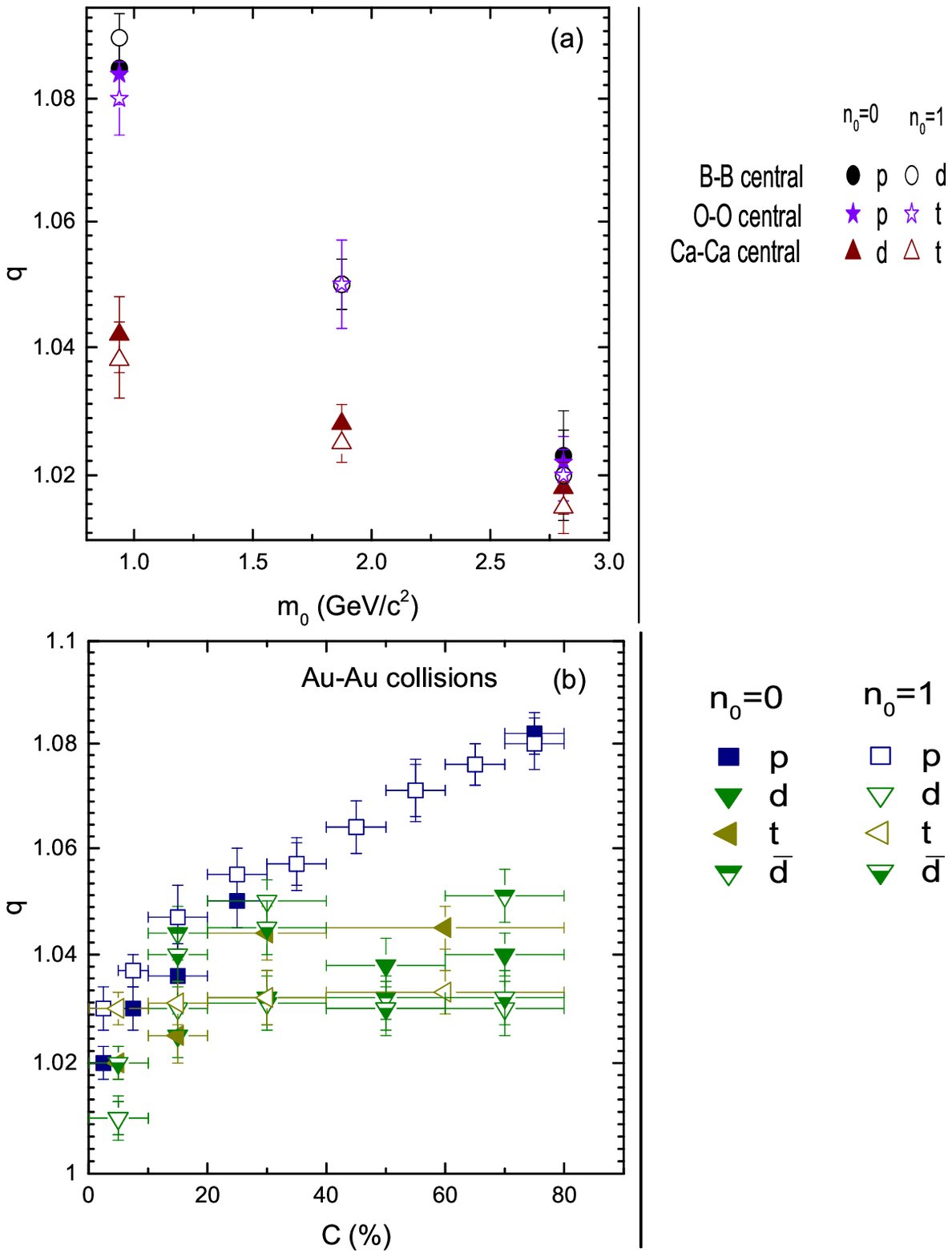}
\end{center}
Fig. 6. Dependence of (a) $q$ on $m_0$ (b) $T_0$ on $m_0$ as well as on centrality.
\end{figure*}

The kinetic freeze-out volume is presented in Fig. 5. Panel (a)
exhibits its dependence on $m_0$ while panel (b) exhibits its
dependence on $m_0$ as well as centrality. $V$ shows dependency on
$m_0$ and it decreases for heavier particles which claim a volume
differential freeze-out scenario and shows the early freeze-out of
heavier particles. From the above discussion, one can assume
different freeze-out surfaces for different particles. From panel
(a) and (b) one can see that $V$ is larger in large systems. In
addition, in panel (b) $V$ is noticed to have a decreasing trend
from central collisions towards the periphery because
hadrons involve in interaction decrease from central to
peripheral collisions depending on their interaction volume. The system with a large number of participants reaches an equilibrium state quickly since there are a larger number of
binary collisions by the re-scattering of partons in central
collisions. The large volume and a large number of participant
nucleons in central collisions could be a hint for the occupation
of super hadronic dense matter. Of course, the present study is
not enough for the study of complete information about the local energy density of super-hadronic matter and the possible phase transition of QGP. However, we will study this in a future
project. $V$ is observed to be larger in the case of $n_0$=1 like
$T_0$ and $\beta_T$. {\bf We would like to clarify that the
resultant data points for $n_0$=0 are not visible because the
values of $n_0$=0 and $n_0$=1 are the same, and they
overlap each other.}

\begin{figure*}[htb!]
\begin{center}
\hskip-0.153cm
\includegraphics[width=15cm]{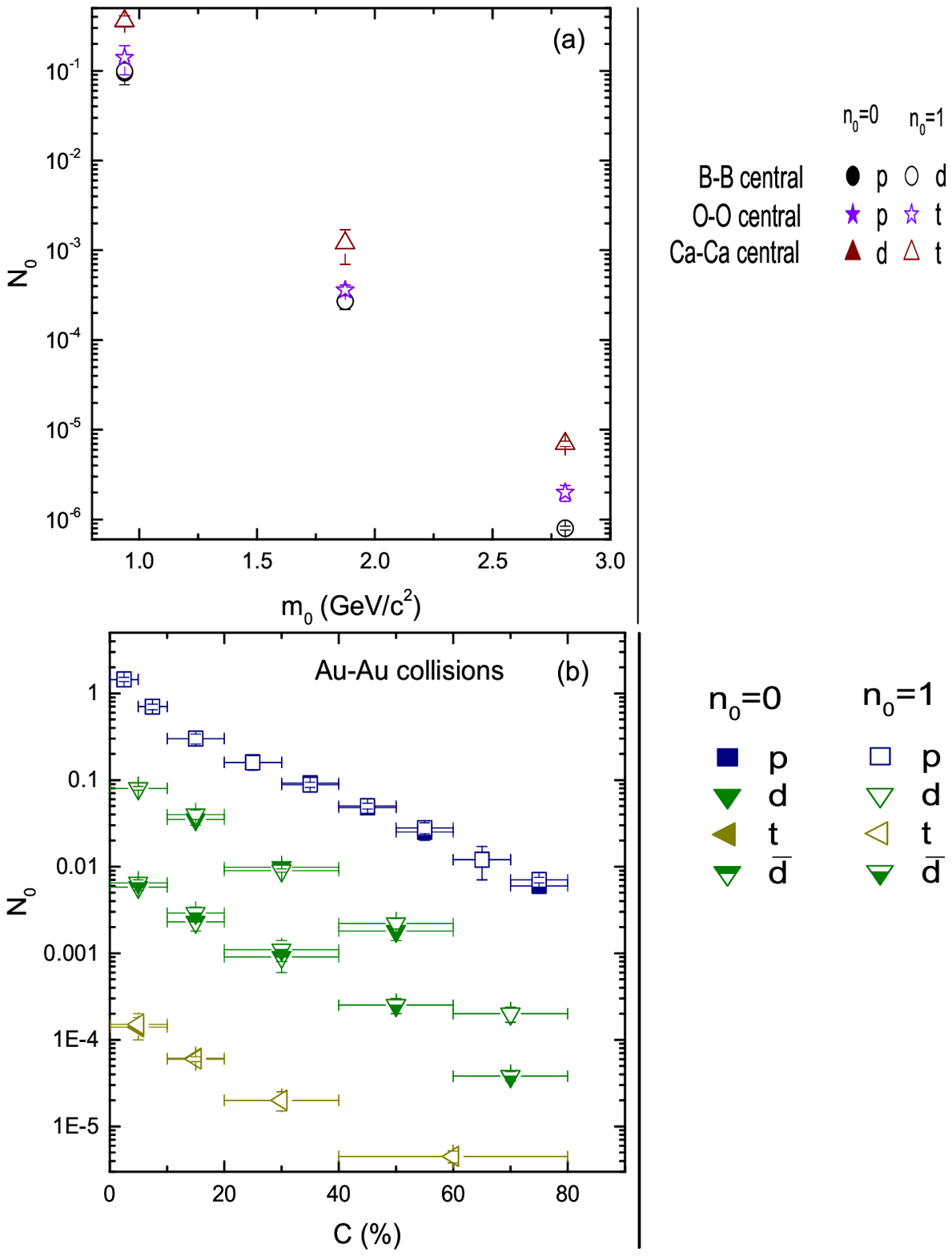}
\end{center}
Fig. 7. Dependence of (a) $\beta_T$ on $m_0$ (b) $T_0$ on $m_0$ as well as on centrality.
\end{figure*}

Fig. 6 is similar to Fig. 5, but it shows the dependence of the
entropy parameter ($q$) on $m_0$ and centrality. $q$ in panel
(b) shows less dependence on $m_0$ in central collisions, but in
panel (a) it is larger for proton and becomes smaller for deuteron
and then triton, even for $Au-Au$ peripheral collisions, $q$ for
proton is the largest. This shows that the production of a proton is
more polygenic than deuteron and triton. In addition, in panel (b)
one can see that $q$ increases from central to a peripheral
collision which indicates a quick approach to an equilibrium state in
central collisions. {\bf  Although $q$ has a non-monotonic
increment from central to peripheral collision in the case of $\bar
d$, but as a whole, it shows an increasing trend from central to
peripheral collisions.}

$q$ is slightly larger in B-B collisions than O-O collisions and
is obviously smaller in Ca-Ca collisions and Au-Au central
collisions which show the dependence of $q$ on the size of the
system. A larger system has a smaller entropy parameter $q$ which
indicates that larger systems have a quick approach to equilibrium
state compared to smaller systems.

The parameter $N_0$ dependence
on $m_0$ and centrality is presented. One can see that the parameter $N_0$ is dependent on $m_0$. The heavier particle has a smaller $N_0$. The parameter $N_0$ is the normalization constant and it shows the multiplicity. The parameter $N_0$ is larger for a larger system which indicates large multiplicity in larger systems, and it decreases from central to peripheral collisions. These results and
natural and understandable.
\\
 {\section{Conclusions}}
 The main observations and conclusions are summarized here.

 a) The transverse momentum spectra of protons, deuterons and tritons are analyzed in Boron-Boron, Oxygen-Oxygen and Calcium-Calcium central collisions and in different centrality intervals in Gold-Gold collisions at 39 GeV using blast wave model with Tsallis statistics with flow profile fixed and linear flow by extracted the kinetic freeze-out temperature ($T_0$), transverse flow velocity ($\beta_T$) and kinetic freeze-out volume ($V$).

 b) $T_0$, $\beta_T$ and $V$ are slightly larger in a linear flow.
$T_0$, $\beta_T$, $V$, entropy parameter ($q$) and the normalization ($N_0$) are mass-dependent. $T_0$, $q$ and $N_0$ increase with increasing the mass of the particle, while $\beta_T$ and $V$ decrease with $m_0$. Furthermore, $T_0$ in central $O-O$ collisions and peripheral $Au-Au$ collisions are close to each other which shows their similar thermodynamic nature.

 c) $T_0$, $\beta_T$, $V$, $q$ and $N_0$ are dependent on the size of the system. They are all larger in large systems except $q$ which is smaller in large systems and indicates a quick approach of the system to equilibrium.

 d) $T_0$ and $\beta_T$ are larger in central collisions and they decrease from central to peripheral collisions due to the decrease of participant
 nucleons towards the periphery which results in less deposition of energy in the system due to less violent collisions in the periphery. $V$ also decreases from central to peripheral collisions due to decreasing the binary collisions by the re-scattering of partons towards the periphery.

 e) $q$ and $N_0$ are centrality dependent. The former increase from the central to the periphery which indicates a quick equilibrium approach of the system in central collisions, while the latter is related to the multiplicity and it decreases from central to peripheral collisions.
\\
\\

{\bf Data availability}

The data used to support the findings of this study are included
within the article and are cited at relevant places within the
text as references.
\\
\\
{\bf Compliance with Ethical Standards}

The authors declare that they are in compliance with ethical
standards regarding the content of this paper.
\\
\\
{\bf Acknowledgements}

The authors would like to thank the support from the National Natural Science Foundation of China (Grant Nos. 11875052, 11575190, and 11135011). Simultaneously, this work is supported in part by the Strategic Priority Research Program of Chinese Academy of Sciences (Grant Nos. XDPB15). The authors would also like to acknowledge the support of Ajman University Internal Research Grant No. DGSR Ref. 2021-IRG-HBS-12.  
\\
\\

\end{multicols}

{\small
\begin{thebibliography}{99}
\setlength{\itemsep}{-1pt}
\bibitem{1}
C.~Shen and U.~Heinz,
Nucl. Phys. News \textbf{25}, no.2, 6-11 (2015)
doi:10.1080/10619127.2015.1006502
[arXiv:1507.01558 [nucl-th]]

\bibitem{2}
I.~Arsene \textit{et al.} [BRAHMS],
Nucl. Phys. A \textbf{757}, 1-27 (2005)
doi:10.1016/j.nuclphysa.2005.02.130
[arXiv:nucl-ex/0410020 [nucl-ex]].

\bibitem{3}
J. Adams et al. (STAR Collaboration), Nucl. Phys. A 757, 102 (2005)

\bibitem{4}
B.~B.~Back \textit{et al.} [PHOBOS],
Nucl. Phys. A \textbf{757}, 28-101 (2005)
doi:10.1016/j.nuclphysa.2005.03.084
[arXiv:nucl-ex/0410022 [nucl-ex]].

\bibitem{5}
A.~Andronic, P.~Braun-Munzinger and J.~Stachel,
Nucl. Phys. A \textbf{772}, 167-199 (2006)
doi:10.1016/j.nuclphysa.2006.03.012
[arXiv:nucl-th/0511071 [nucl-th]]

\bibitem{6}
J.~Cleymans and K.~Redlich,
Phys. Rev. C \textbf{60}, 054908 (1999)
doi:10.1103/PhysRevC.60.054908
[arXiv:nucl-th/9903063 [nucl-th]]

\bibitem{7}
F.~Becattini, J.~Manninen and M.~Gazdzicki,
Phys. Rev. C \textbf{73}, 044905 (2006)
doi:10.1103/PhysRevC.73.044905
[arXiv:hep-ph/0511092 [hep-ph]]

\bibitem{8}
F.~A.~Flor, G.~Olinger and R.~Bellwied,
[arXiv:2109.09843 [nucl-ex]].

\bibitem{9}
R.~Rath, A.~Khuntia and R.~Sahoo,
[arXiv:1905.07959 [hep-ph]].

\bibitem{10}
E.~Laermann and O.~Philipsen,
Ann. Rev. Nucl. Part. Sci. \textbf{53}, 163-198 (2003)
doi:10.1146/annurev.nucl.53.041002.110609
[arXiv:hep-ph/0303042 [hep-ph]].

\bibitem{11}
M. Stephanov, PoS LAT2006, 024 (2006)

\bibitem{12}
K. Rajagopal and F. Wilczek, arXiv hep-ph/0011333 (2000).

\bibitem{13}
J.~Adams \textit{et al.} [STAR],
Phys. Rev. Lett. \textbf{91}, 172302 (2003)
doi:10.1103/PhysRevLett.91.172302
[arXiv:nucl-ex/0305015 [nucl-ex]].

\bibitem{14}
B.~I.~Abelev \textit{et al.} [STAR],
Phys. Rev. Lett. \textbf{97}, 152301 (2006)
doi:10.1103/PhysRevLett.97.152301
[arXiv:nucl-ex/0606003 [nucl-ex]].

\bibitem{15}
J.~Adams \textit{et al.} [STAR],
Phys. Rev. Lett. \textbf{91}, 072304 (2003)
doi:10.1103/PhysRevLett.91.072304
[arXiv:nucl-ex/0306024 [nucl-ex]].

\bibitem{16}
B.~I.~Abelev \textit{et al.} [STAR],
Phys. Rev. C \textbf{77}, 054901 (2008)
doi:10.1103/PhysRevC.77.054901
[arXiv:0801.3466 [nucl-ex]].

\bibitem{17}
B.~I.~Abelev \textit{et al.} [STAR],
Phys. Rev. Lett. \textbf{99}, 112301 (2007)
doi:10.1103/PhysRevLett.99.112301
[arXiv:nucl-ex/0703033 [nucl-ex]].

\bibitem{18}
B.~I.~Abelev \textit{et al.} [STAR],
Phys. Lett. B \textbf{655}, 104-113 (2007)
doi:10.1016/j.physletb.2007.06.035
[arXiv:nucl-ex/0703040 [nucl-ex]].

\bibitem{19}
J.~Adams \textit{et al.} [STAR],
Phys. Rev. Lett. \textbf{95}, 122301 (2005)
doi:10.1103/PhysRevLett.95.122301
[arXiv:nucl-ex/0504022 [nucl-ex]]

\bibitem{20}
M.~Waqas and F.~H.~Liu,
Eur. Phys. J. Plus \textbf{135}, no.2, 147 (2020). doi:10.1140/epjp/s13360-020-00213-1, [arXiv:1911.01709 [hep-ph]]

\bibitem{21}
 M.~Waqas, G.~X.~Peng and F.~H.~Liu,
 J. Phys. G \textbf{48}, no.7, 075108 (2021)

\bibitem{22}
M.~Waqas, F.~H.~Liu, S.~Fakhraddin and M.~A.~Rahim,
Indian J. Phys. \textbf{93}, no.10, 1329-1343 (2019)

\bibitem{23}
M.~Waqas, G.~X.~Peng, F.~H.~Liu and Z.~Wazir,
Sci. Rep. \textbf{11}, no.1, 20252 (2021)

\bibitem{24}
M.~Waqas, F.~H.~Liu and Z.~Wazir, 
Adv. High Energy Phys. \textbf{2020}, 8198126 (2020)

\bibitem{25}
M.~Waqas, F.~H.~Liu, L.~L.~Li and H.~M.~Alfanda,
Nucl. Sci. Tech. \textbf{31}, no.11, 109 (2020)

\bibitem{26}
M.~Waqas and G.~X.~Peng, 
Adv. High Energy Phys. \textbf{2021}, 6674470 (2021)

\bibitem{27}
M.~Waqas, G.~X.~Peng, M.~Ajaz, A.~Haj~Ismail, P.~P.~Yang and Z.~Wazir, 
[arXiv:2112.00975 [hep-ph]].

\bibitem{28}
E. Schnedermann, J. Sollfrank, U.W. Heinz, 
Phys. Rev. C \textbf{48}, 2462 (1993).

\bibitem{29}
M. Ajaz, A.M. Khubrani, M. Waqas, A. Haj Ismail, E.A. Dawi, 
Results in Physics \textbf{35}, (2022) 105433
doi:10.1016/j.rinp.2022.105433

\bibitem{29a}
M.~Waqas and B.~C.~Li, 
Adv. High Energy Phys. \textbf{2020}, 1787183 (2020)

\bibitem{30}
Z.B. Tang, Y.C. Xu, L.J. Ruan et al., 
Phys. Rev. C \textbf{79}, 051901(R) (2009). https://doi.org/10.1103/PhysRevC.79.051901

\bibitem{31}
R. Hagedorn, 
Riv. Nuovo Cimento \textbf{6}(10), 1 (1983). https://doi.org/10.1007/BF02740917

\bibitem{32}
J. Cleymans, D. Worku, 
Eur. Phys. J. A \textbf{48}, 160 (2012). https://doi.org/10.1140/epja/i2012-12160-0

\bibitem{33}
M.~Ajaz, M.~Waqas, G.~X.~Peng, Z.~Yasin, H.~Younis and A.~A.~K.~H.~I.~l,
Eur. Phys. J. Plus \textbf{137}, 52 (2022), doi:10.1140/epjp/s13360-021-02271-5, [arXiv:2112.03187 [hep-ph]]

\bibitem{34}
J. Cleymans, H. Oeschler, K. Redlich et al., 
Phys. Rev. C \textbf{73}, 034905 (2006). https://doi.org/10.1103/PhysRevC.73.034905

\bibitem{35}
R.~P.~Adak, S.~Das, S.~K.~Ghosh, R.~Ray and S.~Samanta,
Phys. Rev. C \textbf{96}, no.1, 014902 (2017)
doi:10.1103/PhysRevC.96.014902
[arXiv:1609.05318 [nucl-th]].

\bibitem{36}
Muhammad Ajaz et al., 
Results in Physics  \textbf{30}, 104790 (20121) 
doi:10.1016/j.rinp.2021.104790

\bibitem{36a}
S. Uddin, J.S. Ahmad, W. Bashir et al., 
J. Phys. G 39, 015012 (2012). https://doi.org/10.1088/0954-3899/39/1/015012

\bibitem{36b}
Pei-Pin Yang, Muhammad Ajaz, Muhammad Waqas, Fu-Hu Liu and Mais Suleymanov 
Journal of Physics G: Nuclear and Particle Physics. \textbf{49}, 055110 (2022) doi:10.1088/1361-6471/ac5d0b 

\bibitem{37}
L.~Adamczyk \textit{et al.} [STAR],
Phys. Rev. C \textbf{96}, no.4, 044904 (2017)
doi:10.1103/PhysRevC.96.044904
[arXiv:1701.07065 [nucl-ex]].

\bibitem{38}
J.~Adams \textit{et al.} [STAR],
Phys. Rev. Lett. \textbf{92}, 112301 (2004)
doi:10.1103/PhysRevLett.92.112301
[arXiv:nucl-ex/0310004 [nucl-ex]].

\bibitem{39}
B.~Abelev \textit{et al.} [ALICE],
Phys. Rev. C \textbf{88}, 044910 (2013)
doi:10.1103/PhysRevC.88.044910
[arXiv:1303.0737 [hep-ex]].

\bibitem{40}
B.~I.~Abelev \textit{et al.} [STAR],
Phys. Rev. C \textbf{79}, 034909 (2009)
doi:10.1103/PhysRevC.79.034909
[arXiv:0808.2041 [nucl-ex]].

\bibitem{41}
B.~B.~Abelev \textit{et al.} [ALICE],
Phys. Lett. B \textbf{728}, 25-38 (2014)
doi:10.1016/j.physletb.2013.11.020
[arXiv:1307.6796 [nucl-ex]]

\bibitem{42}
S.~Acharya \textit{et al.} [ALICE],
Phys. Rev. C \textbf{101}, no.4, 044907 (2020)
doi:10.1103/PhysRevC.101.044907
[arXiv:1910.07678 [nucl-ex]].

\bibitem{43}
O.~Socolowski, Jr., F.~Grassi, Y.~Hama and T.~Kodama,
Phys. Rev. Lett. \textbf{93}, 182301 (2004)
doi:10.1103/PhysRevLett.93.182301
[arXiv:hep-ph/0405181 [hep-ph]].

\bibitem{44}
A.~P.~Mishra, R.~K.~Mohapatra, P.~S.~Saumia and A.~M.~Srivastava,
Phys. Rev. C \textbf{77}, 064902 (2008)
doi:10.1103/PhysRevC.77.064902
[arXiv:0711.1323 [hep-ph]].

\bibitem{45}
W.~Broniowski, M.~Rybczynski and P.~Bozek,
Comput. Phys. Commun. \textbf{180}, 69-83 (2009)
doi:10.1016/j.cpc.2008.07.016
[arXiv:0710.5731 [nucl-th]].

\bibitem{46}
H.~J.~Drescher, S.~Ostapchenko, T.~Pierog and K.~Werner,
Phys. Rev. C \textbf{65}, 054902 (2002)
doi:10.1103/PhysRevC.65.054902
[arXiv:hep-ph/0011219 [hep-ph]].

\bibitem{47}
C. Tsallis, J. Stat. Phys. 52, 479 (1988).

\bibitem{48}
M.~Waqas and F.~H.~Liu,
Indian J. Phys. \textbf{96} (2022) no.4, 1217-1235
doi:10.1007/s12648-021-02058-5
[arXiv:1806.05863 [hep-ph]]     

\bibitem{48a}
M.~Waqas, F.~H.~Liu, R.~Q.~Wang and I.~Siddique,
Eur. Phys. J. A \textbf{56} (2020) no.7, 188
doi:10.1140/epja/s10050-020-00192-y
[arXiv:2007.00825 [hep-ph]].

\bibitem{49}
Y.~L.~Cheng, S.~Zhang and Y.~G.~Ma,
Eur. Phys. J. A \textbf{57}, no.12, 330 (2021)£¬ doi:10.1140/epja/s10050-021-00639-w [arXiv:2112.03520 [nucl-th]]

\bibitem{50}
60. L.~Adamczyk \textit{et al.} [STAR],
''Phys. Rev. C \textbf{96}, no.4, 044904 (2017)doi:10.1103/PhysRevC.96.044904[arXiv:1701.07065 [nucl-ex]].

\bibitem{51}
J.~Adam \textit{et al.} [STAR],
Phys. Rev. C \textbf{99}, no.6, 064905 (2019)
doi:10.1103/PhysRevC.99.064905
[arXiv:1903.11778 [nucl-ex]].

\bibitem{52}
D.~Zhang [STAR],
Nucl. Phys. A \textbf{1005}, 121825 (2021), doi:10.1016/j.nuclphysa.2020.121825
[arXiv:2002.10677 [nucl-ex]].
\
\end{thebibliography}}
\end{document}